\documentclass[%
reprint,
superscriptaddress,
a4paper,
showpacs,preprintnumbers,
amsmath,amssymb,
aps,
prstab,
floatfix,
]{revtex4-1}

\usepackage{graphicx}% Include figure files
\usepackage[caption=false]{subfig}
\usepackage{float}
\usepackage{hyperref}% add hypertext capabilities
\hypersetup{colorlinks=true,citecolor=blue}
\begin{document}
%\preprint{APS/XX-XXX}
\title{Scheme for a 50~TeV muon accumulator}% Force line breaks with \\
%\thanks{This work has been financially supported by the Istituto Nazionale di Fisica Nucleare~(INFN), Italy, Commissione Scientifica Nazionale~5, Ricerca Tecnologica -- Bando n.~20069.}%
\author{O.R. Blanco-Garc\'ia}
\email{oblancog@lnf.infn.it}
\affiliation{INFN-LNF, Via E.~Fermi 40, 00044 Frascati, Rome, Italy}
%% \affiliation{INFN-LNF, Via E.~Fermi 40, 00044 Frascati, Rome, Italy}
%% \author{M. Antonelli}
%% \affiliation{INFN-LNF, Via E.~Fermi 40, 00044 Frascati, Rome, Italy}
%% \author{M. Boscolo}
%% \affiliation{INFN-LNF, Via E.~Fermi 40, 00044 Frascati, Rome, Italy}
%% \author{A. Ciarma}
%% \affiliation{INFN-LNF, Via E.~Fermi 40, 00044 Frascati, Rome, Italy}
%% \author{F. Collamati}
%% \affiliation{INFN-Rome, Piazzale A. Moro 2, 00185 Rome, Italy}
%% \author{S. Guiducci}
%% \affiliation{INFN-LNF, Via E.~Fermi 40, 00044 Frascati, Rome, Italy}
%% \author{P.~Raimondi}
%% \affiliation{ESRF, 71 avenue des Martyrs, 38000 Grenoble, France}
% \altaffiliation[Also at ]{CERN.}%Lines break automatically or can be forced with \\
\date{\today}% It is always \today, today,

\begin{abstract}
  Submitted as input for discussions on a possible low emittance muon accumulator at 50~TeV.\par
  I present a scheme to obtain high quality positive and negative muon beams in terms of low emittance, high charge and high energy using an accumulator at 50~TeV (or tens of TeV) filled with a train of low charge and low emittance muon bunches to be injected in top-up mode on a single bunch circulating in a storage/collider ring at the same energy, where damping from synchrotron radiation merges the bunches with minimal emittance degradation.
\end{abstract}
%\pacs{34.80.Uv, 41.75.-i, 41.85.-p, 41.85.Gy, 41.85.Ja, 78.70.-g}% PACS, the Physics and Astronomy
\maketitle
%\tableofcontents
\section{Introduction}
The construction of a 100~TeV proton--proton collider~(FCC--hh) is currently under study to continue expanding the energy frontier exploration in high energy physics~\cite{fcchh}. Two proton beams, at 50~TeV each, circulate inside two intercepting rings.\par
In this context, it is also worth studying in parallel the possibility of a muon--muon collider in the tens of TeV scale because of the advantages in the kinematics analysis of point-like particle collisions given by leptons and possible access to physics beyond the standard model~\cite{muoncolliders,Zimmermann}.\par
The main difficulty in the realization of a muon collider lies in the limited time (before muons decay) to obtain a highly populated, small emittance and high energy bunch for high luminosity collisions~\cite{boscolo1808}.\par
I would like to present, as input for further discussion, a scheme to produce high quality muon beams (in terms of high charge, high energy and low emittance) obtained from a muon accumulator at 50~TeV filled with a low emittance multi-bunch train to be injected in top-up mode into a single bunch of a main storage/collider ring.\par
Given the intermediate mass of the muon at rest, i.e. 207 times the electron mass and about 1/9 of the proton mass, a muon bunch at 50~TeV in the FCC--hh collider ring would experience total energy loss from synchrotron radiation in a relative small amount of time, which is comparable to one beam life time $\tau$ and very close to one second. This energy loss can be recovered with radio frequency~(RF) cavities in the accelerator that in combination produce a damping effect, well known in electron and positron rings~\cite{damp}.
Therefore, damping should be visible for muons at 50~TeV in a 100~km ring like the FCC--hh design.\par
Profiting from damping in a high energy main ring, we can consider the known method of operation called top-up injection in electron storage rings~\cite{topup}. It consists in the increase of the bunch population by injecting a small additional amount of particles on top of the existing ones in the main ring. In electron rings, particles perform decaying oscillations in the phase space over a time interval of few units of $\tau$ after which they finally reach an equilibrium. The most convenient type of injection for our muon collider/storage ring seems to be on-axis longitudinal injection that consist in the injection of new particles with an offset in time and/or energy with respect to the synchronous bunch. It has the advantage to be twice faster than the transverse damping due to the partition numbers of a typical storage ring.\par
It might be possible that muons show longitudinal damping even at 25~TeV in a ring with a circumference adapted to the reduced life time of muons, thus, lowering the requirements on fast acceleration in previous stages. However, for the moment we leave this as a possibility for further study and concentrate in the 50~TeV scheme to obtain a high intensity bunch at high energy starting from a low charge 22~GeV small emittance muon train given by $e^+e^-$ annihilation of a positron bunch train into fixed targets, studied by the Low EMittance Muon Accelerator~(LEMMA)~\cite{NIM,Antonelli:IPAC16-TUPMY001,Boscolo:IPAC17-WEOBA3,MaricaIPAC2019,eplusringopt} team.\par
The scheme presented in this article could increase the charge of a single muon bunch while preserving or mitigating the growth of the muon beam emittance profiting from synchrotron radiation in the main storage/collider ring.\par
%In the following section we describe the scheme.% to obtain a train of low charge muon bunches stored in a muon accumulator to be injected one by one into the main ring.
\section{Scheme}\label{s:intro}
\begin{figure*}[h]
  \includegraphics[width=0.98\textheight,angle=90]{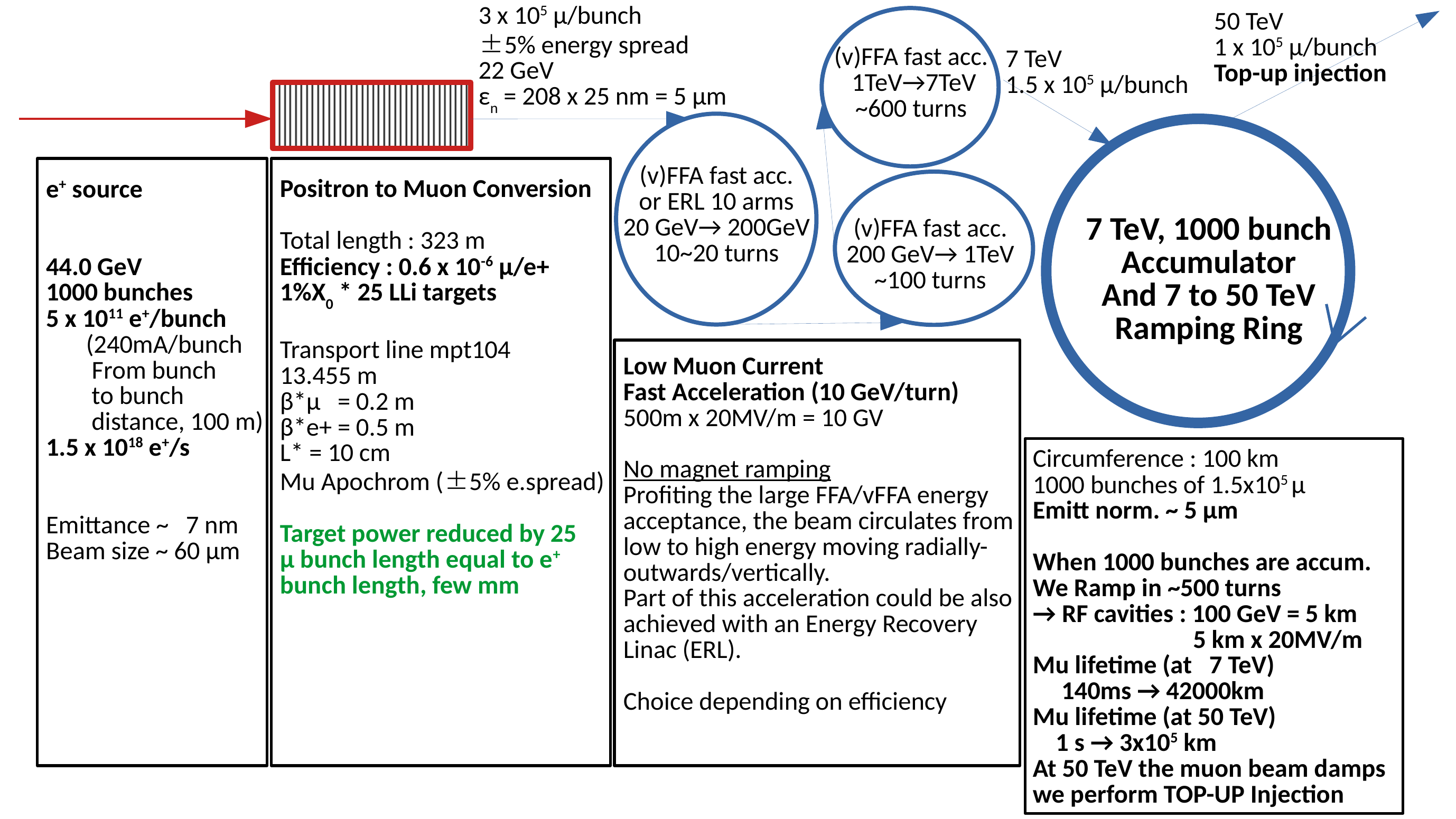}
  \caption{Production and acceleration scheme of a low emittance muon beam for a 50~TeV collider.}\label{f:muacc}
\end{figure*}

The scheme is shown in Fig.~\ref{f:muacc}. It consist in four stages~: a positron source, a positron to muon conversion and transport line, a chain of low muon current fast acceleration rings, and finally a muon accumulator ramping from few TeV to the desired energy (7 to 50~TeV in the scheme). I will describe the stages in the following.
\subsection{Positron source}\label{ss:positronsource}
LEMMA has put forward the idea of a low emittance muon source where muons are produced as secondaries from $e^+e^-$ annihilation of a high energy positron beam, above the energy threshold at 43.7~GeV for muon pair production, impinging on a fixed target.\par
It is convenient to set the positron beam energy to 44~GeV because the energy spread of the produced muon beam will match the acceptance of the next stage. Further details are discussed in Subsection~\ref{ss:muonproduction}.\par
The positron bunch population has been fixed to $5\times10^{11}$, which is a relatively high charge in order to produce as many muons as possible given the small probability of muon pair production, also further discussed in Subsection~\ref{ss:muonproduction}.\par
I fix the positron source requirement to a train of 1000 bunches in order to increase the muon population by a factor 1000 when combining all separate muon bunches into one, and also to be compatible with the positron source requirement studied in~\cite{simone,MaricaIPAC2019,alesini2019positron}.\par
The required positron rate has been estimated to be about $1.5\times10^{18} e^+$/s given a separation of 100 m among neighboring bunches. Notice that the positron bunch separation will be the same muon bunch separation in the train inside the muon accumulator ring.\par
The transverse emittance and beam size is set to be 7~$\pi$~nm and 60~$\mu$m respectively to match the input parameters of the positron to muon conversion stage, discussed in Subsection~\ref{ss:muonproduction}.\par
The energy spread and bunch length are 0.1\% and 3~mm respectively, taken from previous studies of a 6 or 27~km positron ring~\cite{simone}.\par
Finally, I remark that all numbers have been used to clarify and discuss an initial scheme and they could change as required.\par
\subsection{Positron to muon conversion}\label{ss:muonproduction}
The purpose of the positron to muon conversion stage is to produce a low charge and low emittance muon bunch per positron bunch passage.\par
The three particles species ($e^+$ at 44~GeV, and $\mu^+\mu^-$ or simply $\mu$ pairs at 22~GeV) are transported to the final end of the transport line without significative emittance growth of the muon beam, and a small deterioration of positron bunch in terms of emittance and particle population.\par
A detailed description of materials and accelerator optics studies dedicated to this stage can be found in~\cite{blancoprab2020}.\par In this article I have chosen to use a line consisting of 25 liquid Lithium~(LLi) targets of 1\% of a radiation length~$X_0$ connected by a series of quadrupoles for a total length of 323~m. This design is able to produce $0.6\times10^{-6}$ muon pairs per impinging positron, which is factor 10 above a single thin target of 1\%$X_0$ and also distributes the target power deposition.\par
Using the parameters of the positron source shown in Subsection~\ref{ss:positronsource}, we will obtain a bunch population of $3\times10^5$ muon pairs at 22~GeV per positron bunch, with an energy spread of $\pm5$\% in a flat distribution, and with a very small transverse emittance of about 25~$\pi$~nm (equivalent to 5~$\pi$~$\mu$m normalized emittance).\par
The longitudinal emittance will be product of the bunch length of the incoming positron beam by the energy spread from the kinematics of the collision, therefore 3$\times$1~$\pi$~mm~GeV (3~mm $\times$ $0.05\times$22~GeV).\par
For a 22~GeV muon beam the length of the transport line is very short and no significative muon losses are expected.\par
Each one of the 1000 positron bunches will pass one at the time and produce muon pairs that are transported to the acceleration stage. The remaining positrons in the bunch could be spilled or redirected to the previous stage.\par

\subsection{Low Muon Current Fast Acceleration}
Fortunately in the LEMMA scheme  the kinematics of the collision boosts the lifetime by a factor 208 with respect to that of muons at rest, giving an initial time-frame of 0.46~ms of lifetime to accelerate the beam.\par
Muons are already relativistic and therefore I expect acceleration would influence minimally the timing among bunches at different energies.\par
The positive and negative muon beams at 22~GeV will have to be separated and passed through a chain of fast accelerating cavities with the highest possible gradient. I have done an estimation of 500~m of RF cavities at 20~MV/m in order to gain 10~GeV per passage. Although this number could appear high, it is required in order to reduce the number of muon losses due to decay along the acceleration chain. Even higher gradients would be beneficial.\par
There is no definitive structure for the acceleration chain. I have considered three stages as an initial proposal~:
\begin{enumerate}
  \item from 22 to 200~GeV, which is a factor ten in energy gain obtained in ten passages over an RF configuration providing 20~GeV per passage, or twenty turns over an RF configuration at 10~GeV per turn;
  \item from 200~GeV to 1~TeV, which is a factor five in energy gain obtained over approximately 100 turns in a dedicated ring;
  \item from 1~TeV to 7~TeV, which is a factor seven in energy gain obtained over more than 500 turns.
\end{enumerate}
The exact configuration of these acceleration stages needs to be further studied, but, I would like to avoid ramping magnets because of the short lifetime of the muon beam when compared to technological ramping capabilities. Better said, I would prefer to avoid ramping magnets in the KHz range.\par
Several proposals point to technology based in multi-pass Energy Recovery Linacs~(ERL)~\cite{cbeta} and/or Fixed Field Alternating Gradient~(FFA)~\cite{trbojevic,lagrange} ring designs, in particular, vertical FFA (vFFA)~\cite{brooks} to profit from the zero momentum compaction factor and large energy acceptance that could accelerate the beam without the dilution of the longitudinal emittance.\par
%I remark that designs considering vFFA lattices might achieve zero momentum compaction factor along the accelerator chain and therefore it would be possible to consider a smaller positron bunch length, leading to smaller longitudinal muon beam emittance.\par
%In addition, it might be possible to design rings where both positive and muon beam circulate through the same ring and in the same direction as it has been shown in~\ref{bothmuonbeams}.\par
Due to the total number of cycles in the acceleration chain, it is expected to produce a significative muon population loss that has been estimated to be half of the initial bunch, leaving 1.5$\times10^{5}$~muons per bunch.\par

\subsection{Accumulator Ring}
The last section is the Accumulator Ring. It is included last in the scheme because further acceleration in the TeV scale of single muon bunches would require several kilometers of RF cavities and it would be preferable to use a single ring. The ramping frequency is expected to be in the order of few hundreds of Hz, which is a factor 10 above the muon beam life time of 140~ms at 7~TeV that grows up to one second at 50~TeV.\par
The muon life time of 140~ms at 7~TeV also allows for enough time to allocate 1000 individual bunches into separated buckets inside the accumulator with negligible losses due to decay.\par
Once the muon bunch train is in the accumulator, it can be ramped from low to high energy (7~TeV to 50~TeV in the scheme). Then, it could be possible to extract one by one the individual low charge bunches into a single bunch provided the synchronization of the two machines. The most simple case that I can foresee is to have a small difference in circumferences between the accumulator and the main storage/collider ring in order to make coincide once cycle of the bunch in the storage/collider ring with a different bucket of the accumulator every time.\par
We could expect muon losses due to decay given the large number of passages to accelerate~(ramp up) the muon beam train. I have estimated a reduction of the individual bunch to 1.0$\times10^{5}$~muons.\par
Injecting a thousand of these bunches into a storage/collider ring will bring the final bunch charge to about $10^{8}$~muons with a normalized transverse emittance close to 5~$\pi$~$\mu$m, and bunch length and energy spread  dependent on the realization of the acceleration stages.

\section{Conclusion}
I have presented a scheme to produce a high quality muon beam in terms of high particle population, low emittance and energy of 50~TeV (or tens of TeV), using an accumulator ring at the same energy to ramp a low charge, low emittance and low energy train of muon bunches to be injected in top-up mode into a main storage/collider ring.\par
This scheme uses a positron source and a short transport line to produce muon pairs as secondaries from $e^+e^-$ annihilation of a high energy positron beam on fixed targets.\par
Muon pairs are produced at 22~GeV and accelerated in a chain of Fixed Field Accelerator Gradient~(FFA) rings or alternatively Energy Recovery Linacs~(ERL). The number of acceleration stages has been set to three considering the possibility to gain a factor 10 in energy per stage, and the availability of radio-frequency cavities to provide at least 10~GeV per passage to the muon beam.\par
I believe several of the parameters here mentioned (e.g. the RF cavity gradient and the positron to muon conversion efficiency) can be pushed by a factor two or 4, or maybe more, so that the final possible outcome of further studies of this scheme could be very positive.\par
\section{Acknowledgments}
This work has been financially supported by the Istituto Nazionale di Fisica Nucleare~(INFN), Italy, Commissione Scientifica Nazionale~5, Ricerca Tecnologica -- Bando n.~20069, 2019.\par
%The authors would like to thank Helmut Burkhardt for MDISim, used to translate MAD-X optics designs into three-dimensional geometry models for Geant4. We also thank Manuela Boscolo, Mario Antonelli, Susanna Guiducci, Alessandro Variola, Marica Biagini and Francesco Collamati from INFN for useful discussions on the muon beam production. Finally, we thank Pantaleo Raimondi and Simone Liuzzo from ESRF for the original idea of a transport line.\par

%\bibliography{biblio}
\bibliographystyle{unsrt}

\end{document}